\documentclass[aip,pop,preprint]{revtex4-1}
\usepackage{dcolumn}
\usepackage{bm}
\usepackage{amsmath}
\usepackage{tabularx}
\usepackage{algpseudocode}

\usepackage{graphicx}
\usepackage{color}
\usepackage{epsfig}
\usepackage{float}
\usepackage{ulem}

\renewcommand{\emph}[1]{\textit{#1}}
\renewcommand{\Im}[1]{\mathrm{Im}{#1}}

\newcommand{\ave}[1]{\left< #1 \right>}

\renewcommand{\vec}[1]{\boldsymbol{#1}}
\newcommand{\tens}[1]{\vec{\mathsf{#1}}}

\renewcommand{\dh}{\textsc{dh}}

\newcommand{\FT}[1]{\hat{#1}}

\begin{document}


\title{The Barkas Effect in Plasma Transport}


\author{Nathaniel R.~Shaffer}
\affiliation{Los Alamos National Laboratory, Los Alamos, NM, 87545}
\author{Scott D.~Baalrud}
\affiliation{Department of Physics and Astronomy, University of Iowa, Iowa City, IA 52242}


\date{\today}

\begin{abstract}
  Molecular dynamics simulations reveal that a fundamental symmetry of plasma kinetic theory is broken at moderate to strong Coulomb coupling: the collision rate depends on the signs of the colliding charges.
  This symmetry breaking is analogous to the Barkas effect observed in charged-particle stopping experiments and gives rise to significantly enhanced electron-ion collision rates.
  It is expected to affect any neutral plasma with moderate to strong Coulomb coupling such as ultracold neutral plasmas (UNP) and the dense plasmas of ICF and laser-matter interaction experiments.
  The physical mechanism responsible is screening of binary collisions by the correlated plasma medium, which causes an asymmetry in the dynamics of large-angle scattering.
  Because the effect pertains only to close interactions, it is not predicted by traditional transport models based on cut-off Coulomb collisions or linear dielectric response.
  A model for the effective screened interaction potential is presented that is suitable for the coupling strengths achieved in UNP experiments.
  Transport calculations based on this potential and the effective potential kinetic theory agree with simulated relaxation rates and predict that the Barkas effect can cause up to a 70\% increase in the electron-ion collision rate at the conditions of present UNP experiments.
  The influence of the Barkas effect in other transport processes is also considered.
\end{abstract}



\maketitle


\section{Introduction}
\label{sec:intro}

Traditional plasma theory obeys a fundamental symmetry: Coulomb collision rates and associated transport coefficients (such as diffusivity, conductivity, and thermal relaxation) are independent of the signs of the interacting charges.\cite{SpitzerBook}
That is, the values of the transport coefficients do not change if the electrons are instead modeled as positrons.
The origin of this symmetry is the assumption that interactions occur either through binary Coulomb collisions\cite{MontgomeryTidman} or via linear response to weak electrostatic fluctuations.\cite{IchimaruBook}
Using molecular dynamics (MD) simulations, we demonstrate that this symmetry is broken in moderately and strongly coupled plasmas. 
In particular, when $\Gamma_e\gtrsim 0.1$ the electron-ion relaxation rate is significantly larger than the positron-ion relaxation rate at the same conditions.
Here, $\Gamma_e=e^2/a_ek_BT_e$ is the ratio of the Coulomb potential energy of two electrons at the mean inter-electron separation $a_e=(3/4\pi n_e)^{1/3}$ to the average kinetic energy per electron.
The effect is explained using an extension of the Effective Potential Theory (EPT)\cite{BaalrudPRL2013} to charge-neutral plasmas, which quantitatively predicts the simulated relaxation rate. 
It is expected to influence any moderately or strongly coupled neutral plasma, including ultracold neutral plasmas (UNP)\cite{killian_prl_99,ChenPRE2017} and dense plasmas.\cite{ChenPRL2013,ChoNSR2016}

This breaking of charge-sign symmetry is found to be a many-body effect associated with how screening influences low-energy, large-angle collisions. 
Such collisions are rare (and thus negligible) at weak coupling, but they become the dominant type of collision in strongly coupled plasmas.
Despite this, the current leading approaches to collisional transport based on linear response theory do not treat such interactions.\cite{DaligaultPRE2009,VorbergerPRE2010,benedict_pre_12,BenedictPRE2017}
Consequently, such models do not predict the large sign-asymmetry reported here.
In contrast, EPT argues that the essential aspects of many-body screening can be accounted for by treating collisions via the potential of mean force.
This has been shown to provide highly accurate transport rates in one-component plasmas, extending from the weakly coupled limit into the strongly coupled regime (for $\Gamma \lesssim 20$).\cite{BaalrudPRL2013}
However, the theory has not yet been applied to a charge-neutral plasma because a model for the potential of mean force in such a plasma was, until now, not available.
The agreement with MD simulations for charge-neutral plasmas validates key concepts of EPT that are neglected in competing transport models.
Its accuracy also provides compelling evidence that the Barkas effect in collision rates is a large-angle scattering phenomenon.

The general idea that collision rates can depend on the sign of the charges involved is often referred to as ``the Barkas effect''. 
Barkas, Dyer, and Heckman measured that the stopping power of a medium depends on the sign of the projectile's charge,\cite{BarkasPRL1963,AndersenPRL1989} an observation at odds with standard models based on Coulomb scattering or linear dielectric response.\cite{BohrPM1913,BetheADP1930,PinesPR1952,deFerrariisPRA1984}
The Barkas effect is conventionally understood not as a plasma physics phenomenon but as an atomic physics one, where the projectile either excites or polarizes the bound electrons in solids.\cite{JacksonPRB1972,LindhardNIM1976,AshleyPRB1972}
The stopping power of a weakly coupled plasma also exhibits a charge-sign asymmetry when nonlinear polarization of the free electrons is considered, but the effect is small unless the projectile is slow or very highly charged.\cite{PeterPRA1991}
This work presents a new kind of Barkas effect in the macroscopic transport of classical strongly coupled plasmas. 

The proposed effect may be observed in UNP experiments, which create a plasma that is simultaneously neutral, classical, and able to achieve moderate to strong coupling in both species.\cite{killian_prl_99}
This makes UNPs the ideal context in which to test the presence of the Barkas effect in transport processes such as ambipolar diffusion into a vacuum,\cite{KulinPRL2000} electron-ion temperature relaxation,\cite{McQuillenPOP2015} and frictional damping of electron oscillations.\cite{ChenPRE2017}
In particular, the recent measurement of anomalously large electron center-of-mass oscillation damping by Chen et al.\ at $\Gamma_e\sim 0.1-0.35$ (compared with collision models based on repulsive collisions) suggests that the Barkas effect may be responsible.\cite{ChenPRE2017}
Its influence is also expected to extend to electron-ion transport in warm dense plasmas with partially Fermi-degenerate electrons.
The existence of a smooth transition from classical to quantum kinetic behavior with increasing electron degeneracy implies that the effect should be present if the electrons are mildly degenerate.\cite{DaligaultPRL2017}

This paper is organized as follows.
Section~\ref{sec:md} presents the setup, methodology, and results of MD simulations of drift velocity relaxation that show the onset of a Barkas effect.
Section~\ref{sec:theory} reviews the EPT concept for collisional transport and proposes a semi-analytic model for the effective potential appropriate for UNP conditions.
The results of this model are shown to accurately reproduce the relaxation rates in the MD simulations, allowing the Barkas effect to be understood in simple physical terms.
It is also explained why the effect is absent from competing models.
Section~\ref{sec:transport} describes how the Barkas effect is expected to influence other transport processes.
Section~\ref{sec:conclusions} offers some concluding remarks and outlook.

\section{Molecular Dynamics Simulations}
\label{sec:md}

To investigate the Barkas effect at UNP conditions, classical MD simulations of drift velocity relaxation were performed using LAMMPS.\cite{lammps,Plimpton1995}
Two separate cases were considered: the relaxation of electrons or positrons, each on ions.

The essential challenge to simulating electron-ion collisional processes in UNPs is that they spend the entirety of their lifetimes far from thermodynamic equilibrium, with $T_e > T_i$.
  As a consequence, transport coefficients that depend on electron-ion collisions cannot be studied using the techinques of equilibrium MD, i.e., Green-Kubo relations.
  Instead, a non-equilibrium methodology must be adopted, but this introduces the difficulty of deciding which transport process to investigate.
  Temperature relaxation seems like a natural choice, but it is too slow.
  On the long timescale of temperature relaxation, UNPs are subject to other heating and cooling mechanisms (especially three-body recombination heating) that obscure the electron-ion collision physics.\cite{McQuillenPOP2015}
  Velocity relaxation, however, is faster than temperature relaxation by a factor of about $m_i/m_e$.
  This makes it a suitable probe of electron-ion collisional transport (and thus the Barkas effect), since a drift velocity can relax on a timescale that is short compared to both temperature relaxation and three-body recombination.

All simulations described here followed the same basic methodology: a uniform plasma with unequal temperatures was imparted with a small, uniform drift velocity in the electrons or positrons, which subsequently relaxed due to electron-ion or positron-ion collisions.
Section~\ref{sec:md-setup} decribes the simulation setup: the initial and boundary conditions, the interaction potentials, and the choice of time step.
Special attention is given to the question of accurately treating the attractive electron-ion interaction.

In order to connect with the theoretical predictions in Sec.~\ref{sec:theory}, the simulations were conducted such that the relaxation rate could be identified by modeling the drift velocity with a memoryless linear friction law,
\begin{equation}
  \partial_t \vec V_e = -\nu_{ei}\vec V_e
  ,
\end{equation}
assuming a constant relaxation rate $\nu_{ei}$ obtained from an exponential fit to the component of $\vec V_e(t)$ parallel to the perturbation.
This model is justified as long as three criteria are met.
First, the drift kinetic energy had to be negligible compared to the thermal kinetic energy, $\frac{1}{2}m_e V_e^2\ll k_BT_e$.
Second, the transient nonlinear response of the electrons or positrons to the velocity perturbation had to be excluded from the analysis.
Third, the electrons or positrons had to be in a quasi-steady state, meaning that their temperature and radial distribution functions varied little during the relaxation process.
Section~\ref{sec:nemd-method} details how the simulations were choreographed to meet these conditions and allow for the identification of a constant relaxation rate suitable for comparing with theory.

\begin{figure*}
  \centering
  \includegraphics[width=\columnwidth]{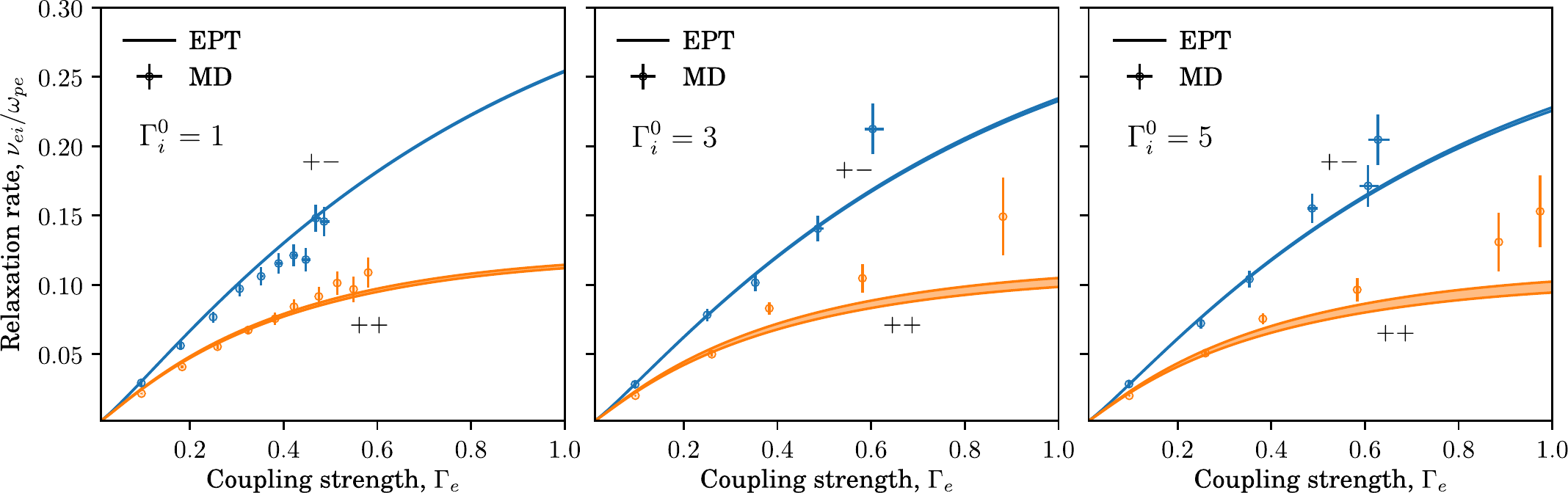}
  \caption{Velocity relaxation rates obtained by MD simulations (symbols) and EPT predictions (curves) for electron-ion plasma (blue) and positron-ion plasma (orange). Vertical error bars indicate $5\sigma$ confidence intervals on the fitted relaxation rate. Horizontal error bars indicate the extremal electron or positron coupling strengths during the fit interval.}
  \label{fig:ept-vs-md}
\end{figure*}
The main results of the simulations are presented in Fig.~\ref{fig:ept-vs-md}, which shows that there exists a Barkas effect in the velocity relaxation rate.
  The fitting procedure for obtaining the MD relaxation rate as well as the meaning of the error bars are described in Appendix~\ref{sec:fit}.
    The theoretical predictions (``EPT'') are those of the model described in Sec.~\ref{sec:theory}.
    The influence of the effect is evident in two consistent trends in the simulation results: the velocity relaxation rate for an electron-ion plasma is faster than that for a positron-ion plasma, and the difference between the two increases the more strongly coupled the electrons/positrons are.
    The onset of the asymmetry is only weakly dependent on the ion coupling strength, but it depends strongly on the electron coupling strength.
    Indeed, the simulation results show that the electrons need only be mildly non-ideal ($\Gamma_e\gtrsim 0.2$) in order for the Barkas effect to influence macroscopic transport.
    This degree of electron coupling has already been surpassed in the UNP experiments on electron center-of-mass oscillation experiments by Chen et al.\cite{ChenPOP2016,ChenPRE2017}
    Thus, it is reasonable to expect that electron-ion transport in present-day UNP experiments may be influenced by the Barkas effect.

  The remainder of this section discusses the simulation methodology in detail.
    An extensive discussion of the motivation, background, and techinical aspects of this work may also be found in Ref.~\onlinecite{ShafferThesis}.

\subsection{Simulation Setup}
\label{sec:md-setup}

All simulations used 5,000 particles of each species.
The ion mass and charge were set to $m_i=1836m_e$ and $Z=1$, respectively, i.e., those of a proton.
The particles were initially distributed uniformly throughout a cubic domain with number densities $n_e=n_i=10^{16}\mathrm{m^{-3}}$.
Periodic boundary conditions were employed to emulate an infinite, homogeneous plasma.
Initial velocities were drawn from distinct Maxwell-Boltzmann distributions for each species so that initially $T_e\ge T_i$.
Since the number density was constant across simulations and throughout their duration, the coupling strength of each species, $s$, was set by the instantaneous value of the kinetic temperature
\begin{equation}
  \label{eq:t-kinetic}
  T_s(t) = \frac{m_s}{3k_BN_s}\sum_{i=1}^{N_s}|\vec v^{(i)}_{s}(t) - \vec V_s(t)|^2
  ,
\end{equation}
where $\vec v_s^{(i)}$ is the velocity of an individual particle of species $s$ and
\begin{equation}
  \label{eq:v-drift}
  \vec V_s(t) = \frac{1}{N_s}\sum_{i=1}^{N_s}\vec v^{(i)}_s(t)
\end{equation}
is the species drift velocity.



The force on each particle was computed by the particle-particle/particle-mesh method using a cutoff radius of $10a_e$ and a $50\times50\times50$ $\vec k$-space mesh.\cite{HockneyEastwood,Plimpton97}
The pairwise interactions for like charges used the Coulomb potential, while the electron-ion pair interactions used the Coulomb potential supplemented by a Gaussian-shaped repulsive core:
\begin{equation}
  \label{eq:pairwise-md}
  \phi_{ss'}(r) = \frac{q_sq_{s'}}{r}\times
  \begin{cases}
    1 & q_sq_{s'} > 0 \\
    1 - e^{-r^2/\alpha^2} & q_sq_{s'} < 0
  \end{cases}
  .
\end{equation}
The repulsive core width, $\alpha$, used in the electron-ion simulations is a purely numerical device to prevent rare encounters between electrons and ions that pass too close to resolve with a fixed time step, leading to poor energy conservation.\cite{kuzmin_prl_02,HockneyEastwood}
A previous study on the thermodynamics of a plasma interacting with the above potential showed that as $\alpha$ decreased, the non-Coulombic features of the electron-ion interaction could be made sufficiently short-ranged as to not affect the macroscopic properties of the plasma.\cite{TiwariPRE2017}
However, since the basic physics of the present velocity relaxation problem are different (non-equilibrium versus equilibrium), it is still necessary to establish convergence of the relevant macroscopic observables with respect to $\alpha$.

\begin{figure}
  \centering
  \includegraphics[width=8.6cm]{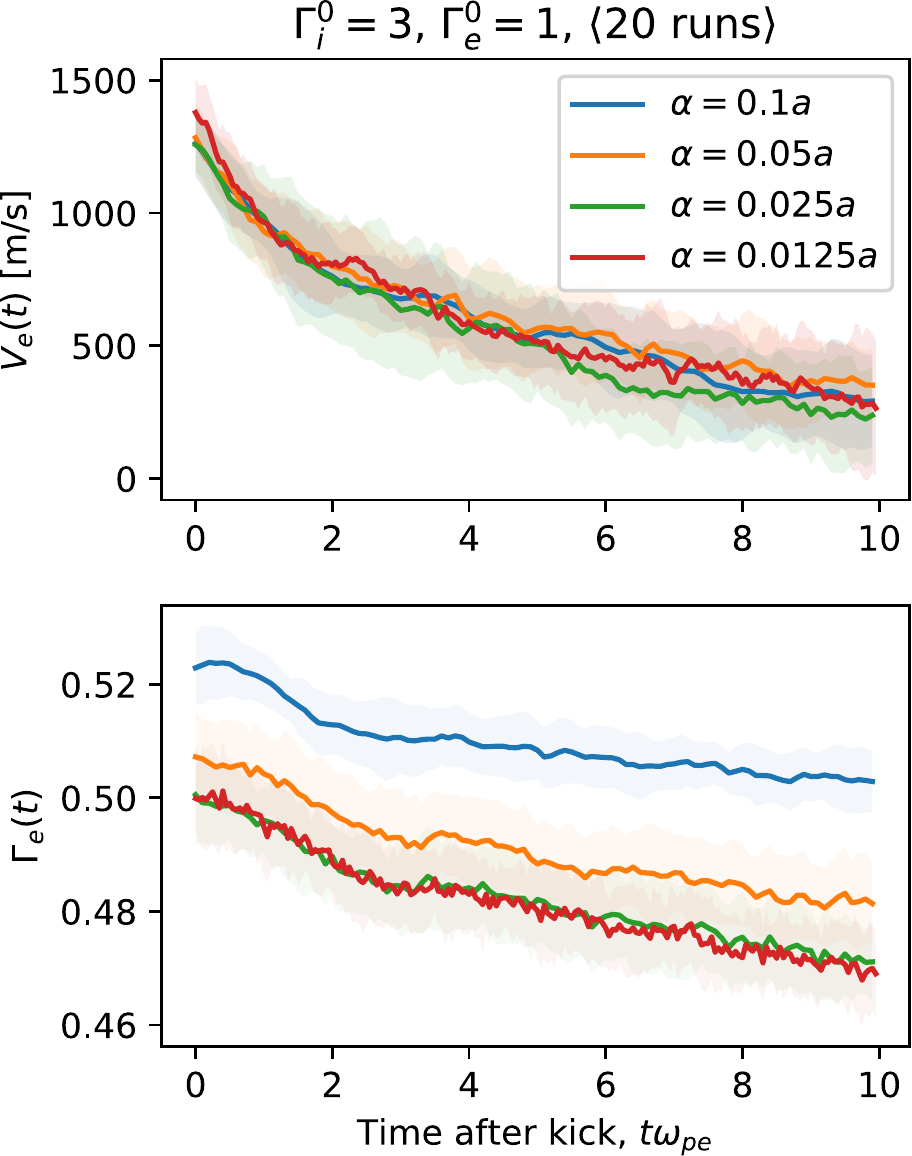}%
  \caption{Convergence of the electron drift velocity (upper) and coupling strength (lower) with respect to the repulsive core width $\alpha$. Curves are the result of averaging over 20 independent simulations, each initialized with $\Gamma_i^0=3$ and $\Gamma_e^0=1$. Shaded regions indicate one sample standard deviation about the mean. The time scale starts at the time after the electron drift velocity was induced, not the start of the simulation.}
  \label{fig:alpha-converge}
\end{figure}

Figure~\ref{fig:alpha-converge} shows convergence tests with initially moderately coupled electrons, $\Gamma_e^0=1$.
Pictured are the instantaneous electron drift velocity and coupling strength as a function of time after the electron drift was induced.
By successively halving $\alpha$, it was found that it could be made sufficiently small not to impact the electron temperature evolution or drift velocity relaxation.
It was found that simulations with initially strongly coupled electrons ($\Gamma_e^0\ge1$) were most sensitive to $\alpha$, but that a repulsive core width of $\alpha=a/80$ was sufficiently small to reach convergence even in these cases.
All simulations of velocity relaxation used $\alpha=a/80$.

\begin{figure}
  \centering
  \includegraphics[width=8.6cm]{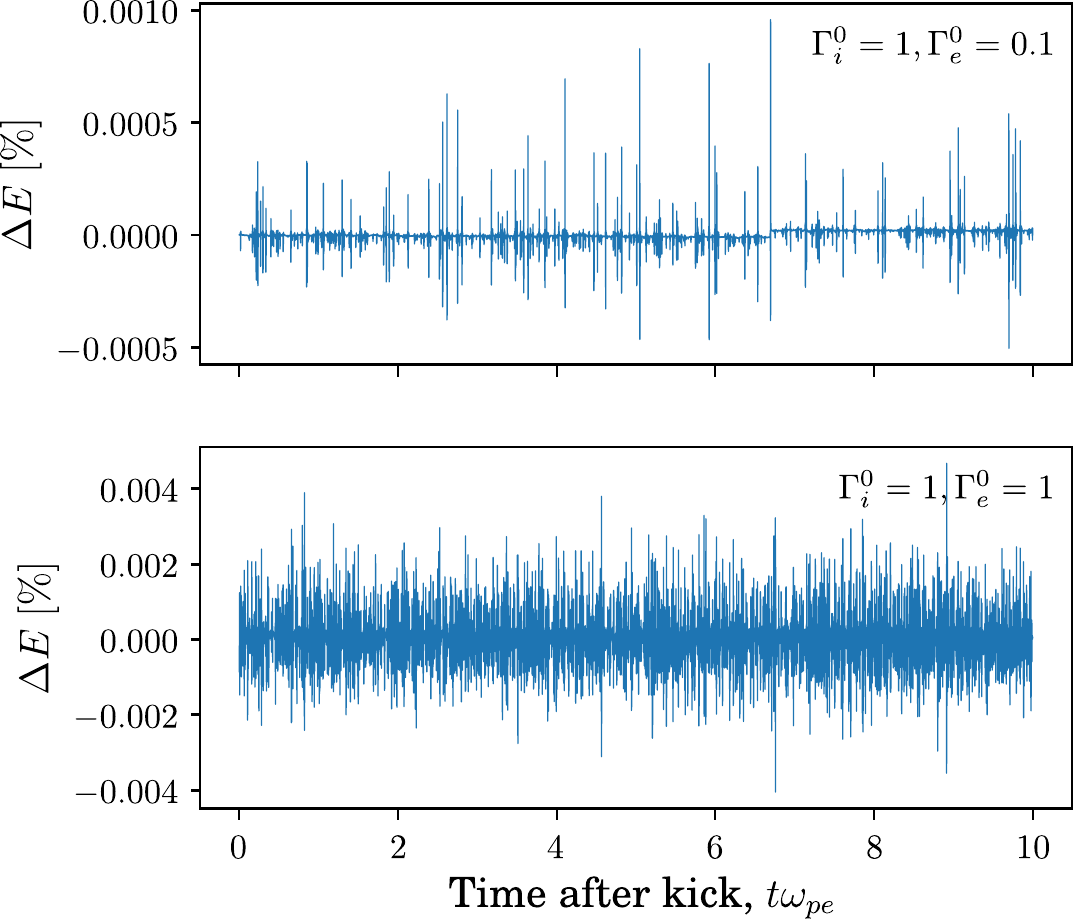}
  \caption{Percent deviation in the total energy from its mean value for an electron-ion simulation with weak initial electron coupling (upper) and moderate initial electron coupling (lower). Only the time after the electron velocity kick is shown because the total energy is discontinuous at the time of the kick.}
  \label{fig:energy-conserve}
\end{figure}

The potential well formed by the repulsive core allowed some electrons to fall into Rydberg-atom-like orbits around ions, but this recombination process occurred slowly enough and the lifetime of bound states was typically short enough that only a small fraction of electrons were bound at any given time.
Even so, the orbital motion of these few bound electrons set the shortest dynamical timescale in the electron-ion simulations.
A rough estimate for this timescale is the orbital period of an electron in circular motion about an ion,
\begin{equation}
  \label{eq:tau-orbit}
  \tau \approx \frac{2\pi}{\omega_{pe}}\sqrt{\frac{3}{Z+1}}(\alpha/a)^{3/2}
  ,
\end{equation}
where the centripetal acceleration is based on the Coulomb force at an orbital radius $\alpha$.
For the repulsive core width $\alpha=a/80$ used in the simulations, the characteristic orbital timescale evaluates to $\tau \approx 0.011\omega_{pe}^{-1}$.
Accordingly, all electron-ion simulations were performed with a time step $\delta t=0.00025\omega_{pe}^{-1}$, corresponding to $\delta t\approx \tau/43$.
Figure~\ref{fig:energy-conserve} shows the variation in total energy of typical electron-ion simulations using this time step, one with initially weakly coupled electrons and one with initially moderately coupled electrons.
There was no appreciable drift in the total energy (which would indicate an overall failure to resolve orbital motion), but sharp fluctuations did occur as a result of occasional close electron-ion interactions.
The frequency and amplitude of these fluctuations increased with the electron coupling, since less energetic electrons were more likely to form deep and/or long-lived bound states.
Bound states did not form in the positron-ion simulations, so a larger time step could be used: $\delta t=0.0025\omega_{pe}^{-1}$ for $\Gamma_e^0\ge 0.1$ and $\delta t=0.00125\omega_{pe}^{-1}$ for $\Gamma_e^0<0.1$.

\subsection{Non-Equilibrium Methodology}
\label{sec:nemd-method}

In order to create conditions conducive to identifying a constant relaxation rate, it was necessary for the simulated plasma to reach a quasi-steady state before inducing a drift velocity.
Physically, this is the condition that the two species are separately in thermal equilibrium with themselves (but not with one another) and that there is minimal exchange of energy between them.
A direct way to test that this is the case is to inspect the evolution of the instantaneous temperature of each species as well as the radial distribution functions (RDFs).
Checking the temperatures diagnoses if there is kinetic energy exchange between species, while the RDFs diagnose if there is potential energy exchange.\cite{HansenMacDonald}

\begin{figure*}
  \centering
  \includegraphics[width=0.5\textwidth]{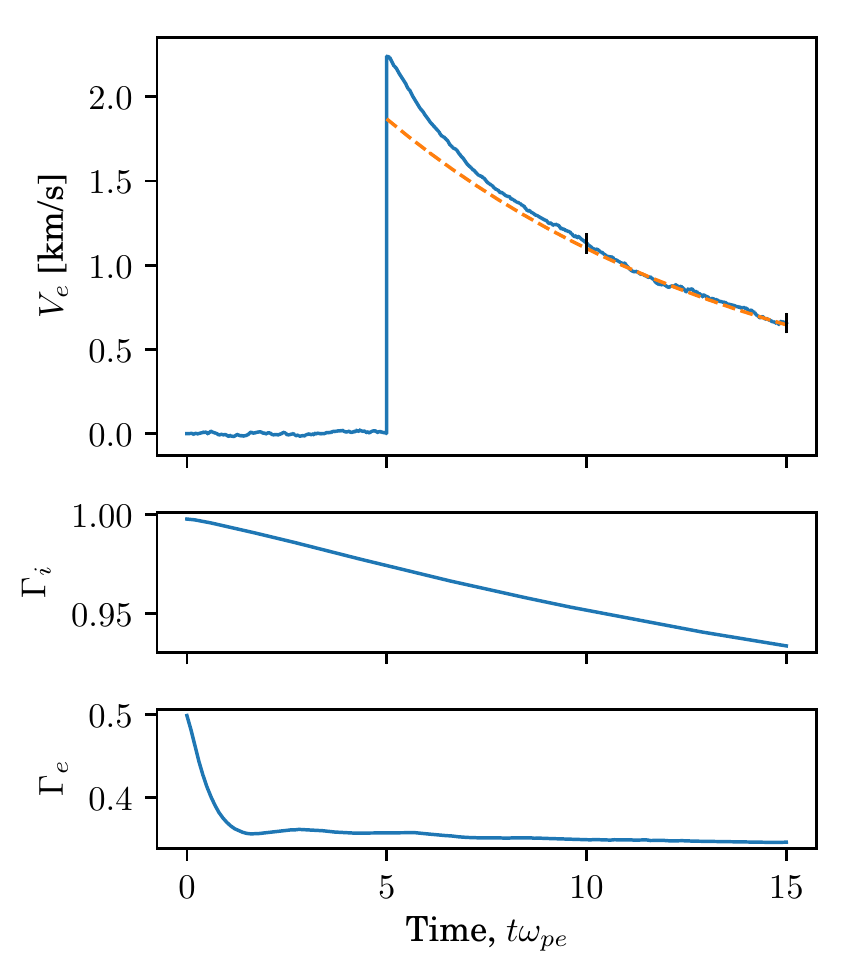}%
  \includegraphics[width=0.5\textwidth]{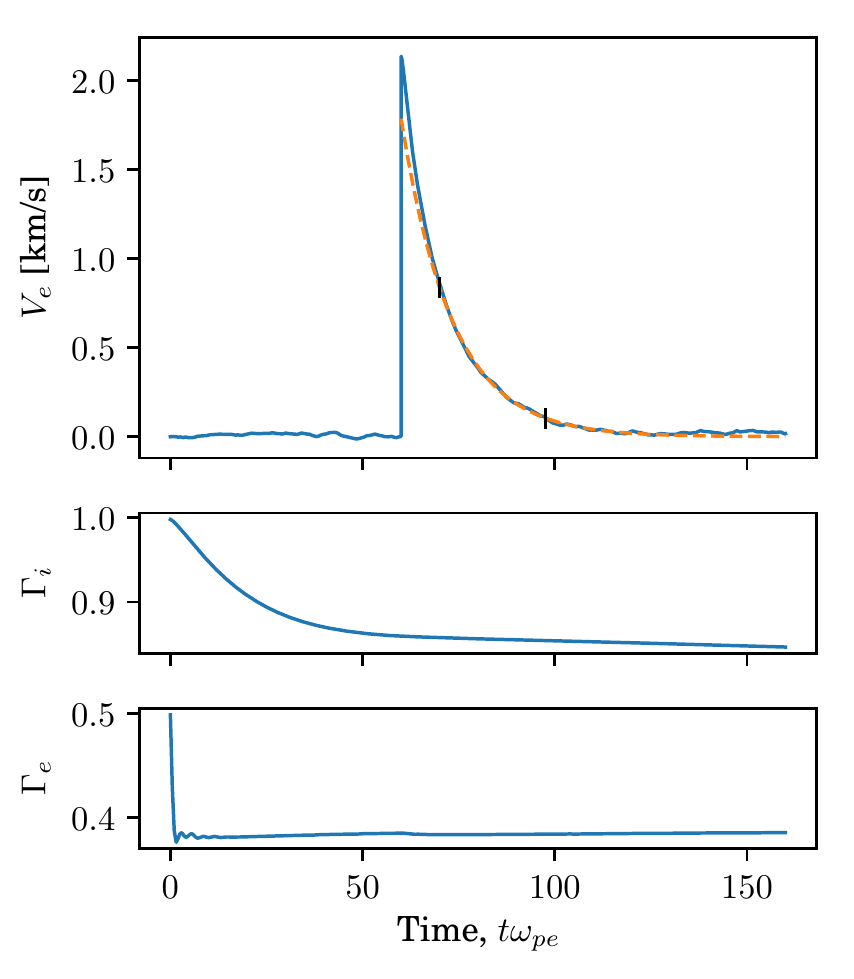}
  \caption{Evolution of drift velocity and coupling strengths for electron-ion (left) and positron-ion (right) simulations, each averaged over 100 independent runs initialized with $\Gamma_i^0=1$ and $\Gamma_e^0=0.5$ (note the differing time scales). The dashed orange line is the fitted model $V_e(t)\propto \exp(-\nu_{ei}t)$. The black vertical bars delimit the time interval used to fit for the relaxation rate. }
  \label{fig:md-example}
\end{figure*}

Figure~\ref{fig:md-example} shows the evolution of the drift velocity and instantaneous coupling strength in two sets of simulations, one of electrons and ions and one of positrons and ions.
In the electron-ion relaxation, it is seen that the ions slowly heated throughout.
This was because the run time of the electron-ion simulations had to be kept shorter than the ion disorder-induced heating (DIH) timescale of $\omega_{pi}^{-1}$ in order to avoid the influence of three-body recombination.
However, the $15\omega_{pe}^{-1}$ run time was long enough for electron DIH to saturate to nearly constant $\Gamma_e$.
In the positron-ion relaxation, the absence of three-body recombination meant that the simulation could have a much longer run time.
Both species were given ample time ($60\omega_{pe}^{-1}$) to finish DIH and reach nearly constant temperatures before the positrons were kicked.
With these long run times, one can see the slight influence of positron-ion temperature relaxation in that $\Gamma_i$ slightly decreased and $\Gamma_e$ slightly increased in the late stages of the simulation.
However, the positron drift velocity decayed away well before there was any appreciable temperature relaxation.

\begin{figure*}
  \centering
  \includegraphics[width=\textwidth]{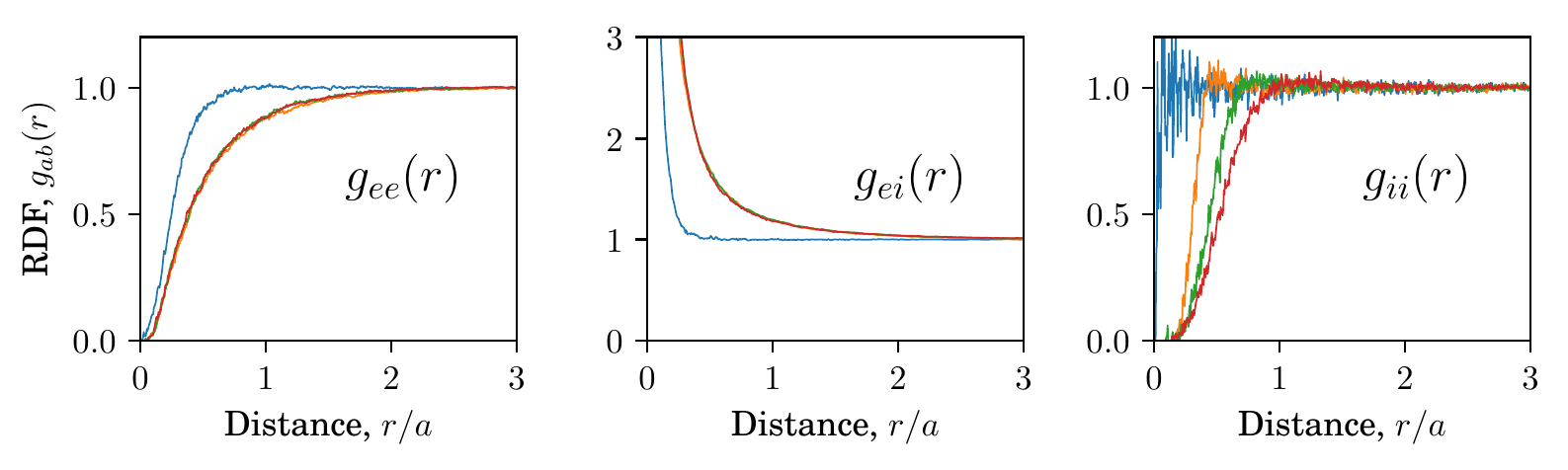}
  \includegraphics[width=\textwidth,trim={0 0 0 1.2cm}]{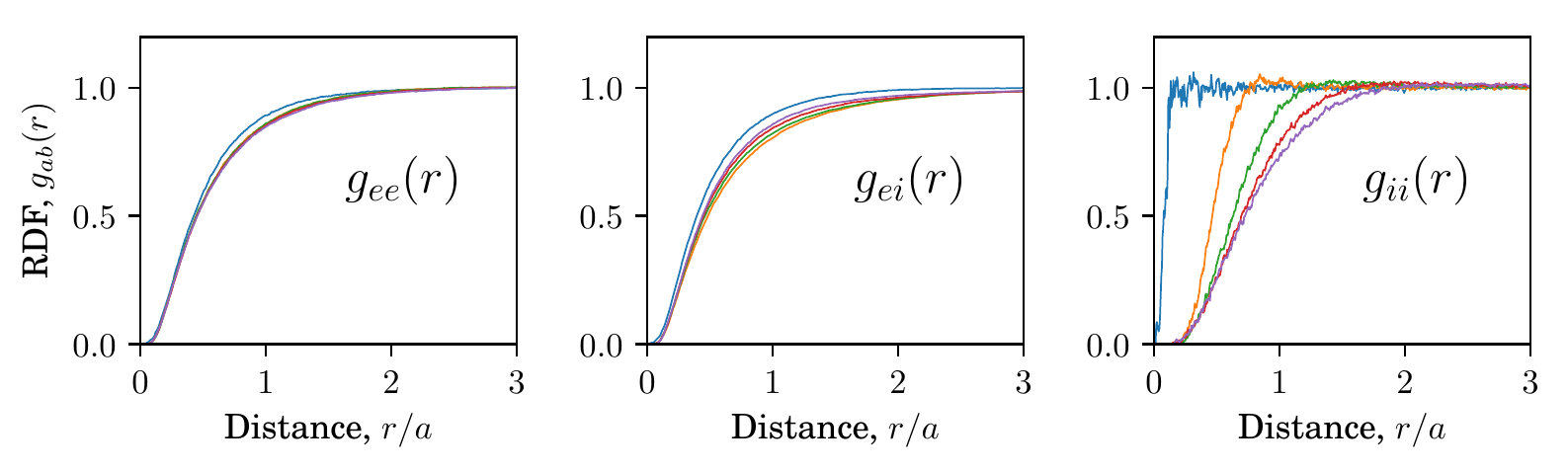}
  \caption{Snapshots of the RDF evolution in an electron-ion plasma (upper) and positron-ion plasma (lower), each initialized with $\Gamma_i^0=1$ and $\Gamma_e^0=0.5$. From left to right: electron-electron/positron-positron, electron-ion/positron-ion, ion-ion. The electron-ion system is shown at $t\omega_{pe}=0,5,10,15$. The positron-ion system is shown at $t\omega_{pe}=0,10,20,30,40$. The time-order of colors is blue $\to$ orange $\to$ green $\to$ red $\to$ purple.}
  \label{fig:rdf-timelapse}
\end{figure*}

The need to kick the positrons at a much later time than the electrons was informed by two observations.
The first was oscillations in the positron coupling strength at early time shown in Fig.~\ref{fig:md-example}, a phenomenon which is analagous to the \emph{ion} kinetic energy oscillations measured in UNPs.\cite{McQuillenPOP2015}
The positron kinetic energy oscillations typically required $10$ to $20\omega_{pe}^{-1}$ to fully decay, depending on their initial coupling strength.
The second and more significant reason for delaying the positron kick was the observation that positrons take a much longer time to form screening clouds around ions than do electrons.
This was observed in the evolution of the RDFs, pictured in Figure~\ref{fig:rdf-timelapse}, which contrasts the evolution of the RDFs in typical electron-ion and positron-ion simulations.
In both cases, $g_{ee}(r)$ quickly reached an essentially static value, signifying that the electrons reached equilibrium with themselves.
Also seen in both cases is that $g_{ii}(r)$ relaxed slowly, since the ions did not yet finish disorder-induced heating to their potential energy minimum on the time scale of the figure.
Where positrons and electrons differed was in $g_{ei}(r)$, which represents the formation of screening clouds around the ions; $g_{ei}(r)$ becomes static when there is no longer appreciable potential energy exchange between the species.
The electron-ion RDF rapidly saturated after DIH, on the order of a few $\omega_{pe}^{-1}$, meaning that quasi-static screening clouds around ions were established.
However, the positron-ion RDF evolved much more slowly, requiring a time on the order of the \emph{ion} plasma period ($\omega_{pi}\approx 43\omega_{pe}^{-1}$) to stabilize.
This was in spite of the fact that positron DIH and kinetic energy oscillations subsided much earlier (recall Fig.~\ref{fig:md-example}, which is for the same data set shown here).

The fact that positrons formed screening clouds much more slowly than electrons has a simple qualitative explanation.
At the start of a simulation, all particles' positions are distributed uniformly randomly.
Electrons rush toward their nearest ion, while positrons rush away from theirs.
Thus, after a short time ($\sim\omega_{pe}^{-1}$), the electrons are already screening the charge of their original nearest ions (which have remained essentially stationary), but the positrons have been redirected to some other ion's vicinity.
The positrons continue to be jostled around by ions until they have settled into the interstitial spaces between ions.
This requires a time on the order of $\omega_{pi}^{-1}$ because where there are clusters of a few ions (as will be frequently be the case for an initially random configuration), it is unlikely that a positron will be able to occupy the space between them until enough time has passed for the ions to repel one another.
Only once they have moved apart and their potential barrier lowered can a positron enter the space.

After the initial waiting period ($5\omega_{pe}^{-1}$ for electrons, $60\omega_{pe}^{-1}$ for positrons) the drift velocity was introduced.
An impulse was delivered to the electrons or positrons over a single time step in order to induce a drift velocity small in magnitude compared to the  instantaneous thermal speed, $V_e(t_{\mathrm{kick}})=0.1v_{Te}(t_{\mathrm{kick}})$, where $v_{Ts}(t)=\sqrt{2k_BT_s(t)/m_s}$.
The plasma was then evolved free from external perturbation to allow the drift velocity to relax.
For the electrons, simulations were carried out an additional $10\omega_{pe}^{-1}$, which was long enough for their drift velocity to appreciably relax but not so long that a significant population of bound electron-ion pairs could form.
Otherwise, inelastic electron-``Rydberg'' collisions would contribute an unknown but potentially significant amount to the total damping rate, which would impede comparison with theory.
The positrons were allowed to relax for $100\omega_{pe}^{-1}$ after the kick, since recombination was not a concern.

For later comparison with theoretical models, it was important to ensure the velocity distribution functions, $f_s(\vec v,t)$, did not significantly deviate from a flow-shifted Maxwellian distribution throughout the relaxation process.
A simple test of this assumption was done by inspecting the time-evolution of the normalized skewness and excess kurtosis of the reduced distributions, $\bar f(v_z,t)=\iint f_s(\vec v,t) dv_xdv_y$.
The only non-Maxwellian feature observed was a slight positive excess kurtosis that would develop during the DIH of either species.
As a stringent cross-check, the kinetic pressure tensor $\tens p_s(t)=\ave{m_s[\vec v - \vec V_s(t)][\vec v-\vec V_s(t)]}$ and heat flux $\vec q_s(t) = \ave{\frac{1}{2}m_s[\vec v-\vec V_s(t)]|\vec v-\vec V_s(t)|^2}$ were computed from the full distribution functions of an electron-ion simulation with $\Gamma_i^0=1$ and $\Gamma_e^0=0.1$.
From these, it was confirmed that throughout the simulation, equipartition of kinetic energy was satisfied ($|p_{ii,s}/n_sk_BT_s|<1.05$), viscous stresses were negligible ($|p_{i\ne j,s}/n_sk_BT_s|<0.02$), and heat flux was negligible ($|q_{i,s}/n_sk_BT_sv_{Ts}|<0.05$).
Thus, to a good approximation, the MD relaxation process could be modeled assuming flow-shifted Maxwellian distributions throughout.

\section{Theory}
\label{sec:theory}


\subsection{Effective Potential Theory}
\label{sec:ept}

EPT is a plasma kinetic theory based on a Boltzmann-like collision operator in which transport occurs through binary interactions between particles, but where many-body effects are included by modeling those interactions as occurring via the potential of mean force.\cite{BaalrudPRL2013}
In a collision between particles of species $s$ and $s'$ with impact parameter $b$ and incident relative speed $u = |\vec v_s-\vec v_{s'}|$, the angle of deflection is
\begin{equation}
  \label{eq:theta}
  \theta_{ss'}(b,u) = \pi - 2b \int_{r_{\min}}^\infty\frac{dr}{r^2\sqrt{1 - \Phi_{ss'}(r,b,u)}}
\end{equation}
where
\begin{equation}
  \label{eq:scat-pot}
  \Phi_{ss'}(r,b,u) = \frac{b^2}{r^2} + \frac{2\phi_{ss'}(r)}{m_{ss'}u^2}
\end{equation}
is the dimensionless scattering potential containing both centrifugal repulsion and the inter-particle potential.
Above, $r_{\min}$ is the distance of closest approach (largest solution to $\Phi_{ss'}=1$ at fixed $b$ and $u$) and $m_{ss'}=m_sm_{s'}/(m_s+m_{s'})$ is the reduced mass.\cite{Mechanics}
Transport coefficients may be expressed in terms of the functions
\begin{equation}
  \label{eq:xi-lk}
  \Xi_{ss'}^{(l,k)} = \frac{1}{2}\int_0^\infty  \frac{\sigma_{ss'}^{(l)}(\xi)}{\sigma_{0,ss'}} \xi^{2k+3}e^{-\xi^2}d\xi
  ,
\end{equation}
which are generalizations of the ``Coulomb logarithm'' appearing in weakly coupled theory.\cite{BaalrudPRL2013}
In Eq.~\eqref{eq:xi-lk},
\begin{equation}
  \label{eq:sigma}
  \sigma_{ss'}^{(l)}(v) = \int_0^\infty[1-\cos^l\theta_{ss'}(v,b)]2\pi b\,db
\end{equation}
is the order-$l$ momentum-transfer cross-section,\cite{FerzigerKaper} $\xi=u/\bar v_{ss'}$ is a dimensionless relative speed, $\bar v_{ss'}=\sqrt{v_{Ts}^2+v_{Ts'}^2}$ is the thermal speed associated with the distribution of relative velocities, and $\sigma_{0,ss'} = \pi (q_sq_{s'}/m_{ss'}\bar v_{ss'}^2)^2$ is a reference cross-section.

The premise of EPT is that the appropriate choice of the interaction potential is not the Coulomb interaction, but rather the potential of mean force, $w_{ss'}(r)$.\cite{BaalrudPRL2013}
The potential of mean force is defined as the potential one obtains by holding two particles at fixed positions and ensemble-averaging over the positions of the other particles.
Usually, it is defined only for statistical ensembles with a single temperature, in which case it is related to the RDF by $w_{ss'}(r) = -k_BT\ln g_{ss'}(r)$.\cite{HillStatMech}
For two-temperature plasmas with weak electron-ion coupling, the Boercker-More ensemble may be used to define the potentials of mean force, e.g., $w_{ei}(r)=-k_BT_e\ln g_{ei}(r)$.\cite{BoerckerPRA1987}

In the limit of weak coupling, the potential of mean force is the Debye-H\"uckel potential, $w_{ei}^\dh(r) = \pm \frac{Ze^2}{r}e^{-r/\lambda_D}$.\cite{HansenMacDonald}
Binary scattering through a DH potential does result in a momentum-transfer cross section that depends on whether the collision is attractive ($-$) or repulsive ($+$).\cite{KhrapakPRL2003,BaalrudPOP2012}
However, this asymmetry is present only in low-velocity, large-angle collisions such that $\frac{1}{2}m_{ei}u^2\ll Ze^2/\lambda_D$, which are infrequent in a weakly coupled plasma.
The resulting order-unity corrections to the Coulomb logarithm are negligible at weak coupling.\cite{LiboffPF1959,BaalrudPOP2012}
One may expect a larger effect at increased coupling, but the DH potential is no longer valid in this regime, necessitating a more accurate model for the potential of mean force.



\subsection{Model for the Potential of Mean Force}
\label{sec:model}

For plasmas in thermal equilibrium (i.e., those characterized by a single temperature $T$), an accurate potential of mean force may be obtained from the RDFs that solve the Ornstein-Zernike (OZ) relations
\begin{equation}
  \label{eq:oz}
  \FT h_{ss'}(k) = \FT c_{ss'}(k) + \sum_{\sigma} n_{\sigma}\FT c_{s\sigma}(k)\FT h_{\sigma s'}(k)
  ,
\end{equation}
paired with the hypernetted chain (HNC) closure
\begin{equation}
  \label{eq:hnc-eq}
  g_{ss'}(r) = \exp\left[ -\frac{\phi_{ss'}(r)}{k_BT} + h_{ss'}(r) - c_{ss'}(r) \right]
  ,
\end{equation}
where $h_{ss'}(r)=g_{ss'}(r)-1$ is the total correlation function, $c_{ss'}(r)$ is the direct correlation function, hats denote Fourier transforms, and the sum in Eq.~\eqref{eq:oz} runs over all species labels.
There is no exact extension of the theory to two-temperature plasmas, but the proposal by Seuferling, Vogel, and Toepffer (SVT)
\begin{subequations}
  \begin{equation}
    \label{eq:hnc}
    g_{ss'}(r) = \exp\left[ -\frac{\phi_{ss'}(r)}{k_BT_{ss'}} + h_{ss'}(r) - c_{ss'}(r) \right]
  \end{equation}
  \begin{equation}
    \label{eq:svt}
    \FT h_{ss'} = \FT c_{ss'}
    + \frac{T_{ss'}}{m_{ss'}}  \sum_{\sigma}n_{\sigma}\left(
      \frac{T_{s\sigma}}{m_s}\FT c_{s\sigma}\FT h_{\sigma s'}
      +\frac{T_{\sigma s'}}{m_{s'}}\FT h_{s\sigma}\FT c_{\sigma s'}
    \right)
  \end{equation}
  \begin{equation}
    \label{eq:cross-temp}
    T_{ss'} = \frac{m_{s'}T_s+m_sT_{s'}}{m_s+m_{s'}}
  \end{equation}
\end{subequations}
was recently shown to yield RDFs and static structure factors that compare favorably with MD simulations of two-temperature positron-ion plasma.\cite{SeuferlingPRA1989,ShafferPOP2017}

The derivation of the SVT equations relies on an assumption that the $N$-body phase-space distribution may be taken in its static limit, that its velocity and position dependence may be decoupled, and that the velocity dependence may be further decoupled into a product of Maxwellians for each species with appropriate temperatures for each.
It is also assumed that any relative drift between the species is negligible.
These assumptions are consistent with the quasi-steady state constructed in the MD simulations.

It is difficult to compute the RDFs of an electron-ion plasma from either the SVT or OZ equations.
The reason is uncontrolled numerical errors that arise from using the attractive Coulomb interaction in the HNC closure.
Since the $r\to0$ behavior of the RDFs is controlled by the Coulomb interaction, $g_{ei}(r)\sim\exp{(1/r)}$ for attractive interactions.
By rearranging the HNC closure as $c_{ss'} = -\phi_{ss'}/k_BT_{ss'} + g_{ss'} -\ln g_{ss'} - 1$, it is evident that $c_{ei}(r)$ also diverges exponentially as $r\to0$.
On the one hand this behavior seems physically reasonable, since the potential of mean force should recover the bare interaction at close separation.
On the other hand, since $g_{ei}(r)$ and $c_{ei}(r)$ each have a non-integrable singularity at $r=0$, they do not have Fourier transforms.
Consequently, the SVT or OZ equations cannot be solved in the usual way.\cite{SpringerJCP1973,NgJCP1974}

An attempt was made to circumvent these numerical difficulties by artificially softening the electron-ion interaction
\begin{equation}
  \label{eq:phi-soft}
  \phi_{ei}(r) \to -\frac{Ze^2}{r}(1 - e^{-r/\alpha})
  ,
\end{equation}
which takes the value $\phi_{ei}(0) = -Ze^2/\alpha$ at zero separation.
This introduces a new parametric dependence on the softening length scale $\alpha$, and it is the $\alpha\to0$ limit which is relevant to UNPs.
To reach the lowest numerically feasible value of $\alpha$, a sequence of HNC calculations at fixed $\Gamma_i$ and $\Gamma_e$ were performed with successively decreasing $\alpha$.
It was found that the iterative solution to the HNC-SVT equations could only be made to converge when $\alpha/a_e\gtrsim \Gamma_e/10$.
At the moderate electron coupling strengths where the Barkas effect was observed in MD, this was not small enough to extrapolate to an accurate $\alpha\to0$ limit.
Details of this attempt and its failure to describe the MD simulation results are given in Appendix~\ref{sec:alpha-extrap}.

A more fruitful approach is to exploit the fact that the electron coupling strength is not very high in UNPs, which enables a connection between the SVT equations and the screened one-component plasma model.\cite{ShafferPOP2017}
The SVT equations in the limit  $m_e/m_i\to0$ can be rearranged in terms of the partial static structure factors $S_{ss'}(k) = \delta_{ss'} + \sqrt{n_sn_{s'}}\FT h_{ss'}(k)$ as
\begin{subequations}
  \begin{align}
    & S_{ii} = \frac{1}{1 - n_i\FT c_{\mathrm{scr}}} \\
    & S_{ei} = \frac{\sqrt{n_en_i}\FT c_{ei}}{1 - n_e\FT c_{ee}} S_{ii} \\
    & S_{ee} = \frac{1}{1 - n_e\FT c_{ee}} + \frac{n_en_i\FT c_{ei}^2}{(1 - n_e\FT c_{ee})^2} S_{ii} \\
    & \FT c_{\mathrm{scr}} = \FT c_{ii} + \frac{T_e}{T_i}\frac{n_e\FT c_{ei}^2}{1-n_i\FT c_{ii}}
      .
  \end{align}
\end{subequations}
Taking $n_e=Zn_i$ and treating the electrons (or positrons) in the Debye-H\"uckel approximation amounts to setting
\begin{subequations}
  \label{eq:ei-ee-dh}
  \begin{align}
    & \FT c_{ei}(k) \approx -\frac{\FT \phi_{ei}(k)}{k_B T_{ei}} = \pm\frac{1}{\sqrt{n_en_i}} \frac{\sqrt Z}{\lambda_{De}^2k^2}
    \\
    & \FT c_{ee}(k) \approx -\frac{\FT \phi_{ee}(k)}{k_B T_e} = -\frac{1}{n_e}\frac{1}{\lambda_{De}^2k^2}
      ,
  \end{align}
\end{subequations}
where plus or minus signs refer to electrons and positrons, respectively.
One then finds that the electron-ion structure factor simplifies to
\begin{equation}
  \label{eq:Sei}
  S_{ei}(k) \approx \pm\frac{\sqrt{Z}S_{ii}(k)}{1+\lambda_{De}^2k^2}
  ,
\end{equation}
and the ion-ion structure factor becomes that of a one-component plasma with coupling strength $\Gamma=\Gamma_i$ and screening parameter $\kappa=a_i/\lambda_{De}=Z^{1/3}\sqrt{3\Gamma_e}$.\cite{ShafferPOP2017}
Note that in this treatment, only the electrons (or positrons) are treated in the Debye-H\"uckel approximation; no assumption on the ion coupling strength is made.

In the present approximation, the structure factors are in fact the same as those derived by Boercker and More from their two-temperature statistical ensemble.\cite{BoerckerPRA1987}
The HNC closure, Eq.~\eqref{eq:hnc}, may then used to compute an approximate electron-ion potential of mean force, $w_{ei}(r) = \phi_{ei}(r) - k_B T_e[ h_{ei}(r) - c_{ei}(r)]$ which is consistent with its statistical definition in terms of the Boercker-More ensemble.
In this approximation, 
\begin{align}
  w_{ei}(r)
  &\approx -k_B T_e \frac{1}{8\pi^3}\int \FT h_{ei}(k) e^{i\vec k\cdot\vec r}d\vec k
  \\
  &\approx \mp\frac{Ze^2}{r}\cdot \frac{2}{\pi}{\lambda_{De}^2}\int_{0}^\infty \frac{k \sin(kr)}{1 + \lambda_{De}^2k^2}S_{ii}(k)\,dk  \label{eq:dh-pmf}
    ,
\end{align}
meaning the electron-ion potential of mean force may be computed with only the knowledge of the structure factor of an appropriately parameterized screened OCP.

\begin{figure}
  \centering
  \includegraphics[width=8.6cm]{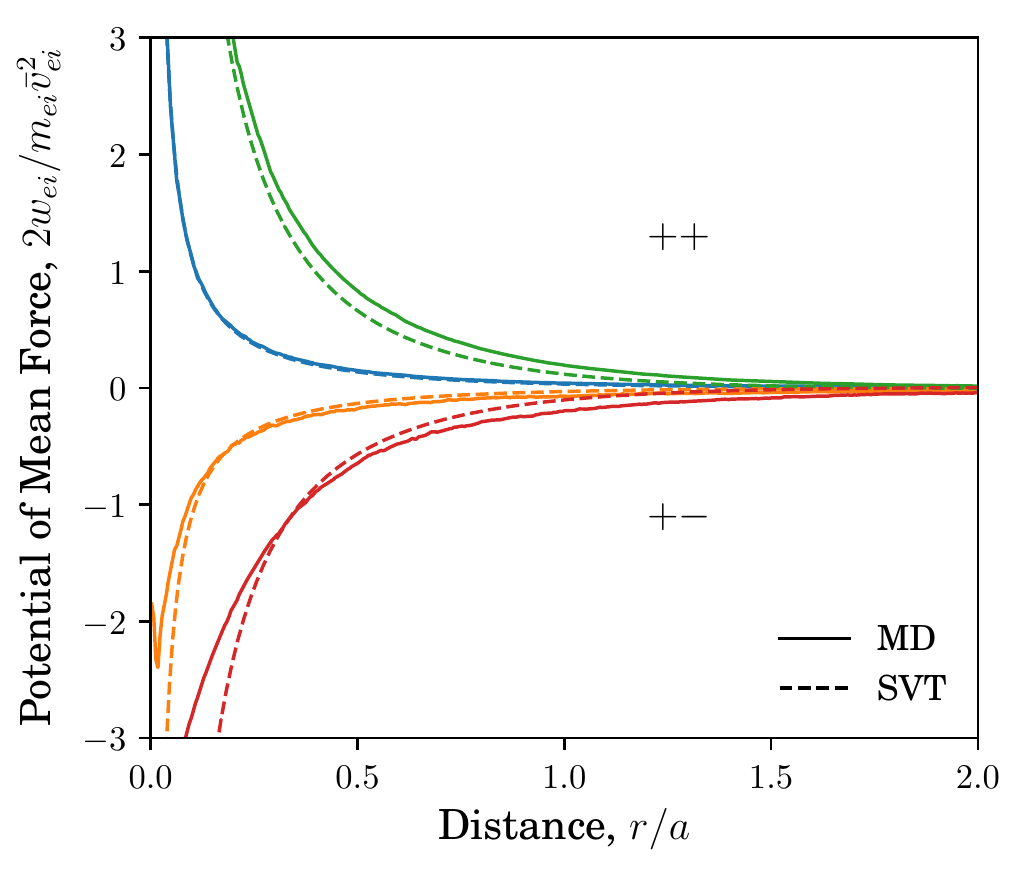}
  \caption{Potentials of mean force obtained from MD simulations (solid lines) compared with Eq.~\eqref{eq:dh-pmf} (dashed lines).
    The upper curves ($++$) are for positron-ion plasmas.
    Blue: $\Gamma_i=0.83$, $\Gamma_e=0.10$;
    green: $\Gamma_i=0.92$, $\Gamma_e=0.58$.
    The lower curves ($+-$) are for electron-ion plasmas.
    Orange: $\Gamma_i=0.95$, $\Gamma_e=0.10$;
    red: $\Gamma_i=0.96$, $\Gamma_e=0.49$.
  }
  \label{fig:wei-svt-vs-md}
\end{figure}

The approximations made in Eq.~\eqref{eq:ei-ee-dh} are appropriate when the electrons are not strongly coupled.
This is the case in all UNP experiments done to date, where $\Gamma_e\le0.35$,\cite{ChenPRE2017} as well as in the majority of the MD simulations described in Sec.~\ref{sec:md}.
The accuracy of Eq.~\eqref{eq:dh-pmf} for the potential of mean force has been verified both by comparison with fully two-component SVT solutions for positron-ion plasmas,\cite{ShafferPOP2017} as well as by comparison with MD simulations, shown in Fig.~\ref{fig:wei-svt-vs-md}.

An advantageous side-effect of using this approximation is that Eq.~\eqref{eq:dh-pmf} differs between electrons and positrons only in overall sign of the potential.
In other words, even though positrons and electrons have different charge, in the limit that they are not too strongly coupled, they screen the ions identically.
This means that whatever Barkas effect is exhibited in the EPT results that follow from this approximation arise solely due to this leading sign of the effective potential.
This is an important point for identifying the physical mechanism responsible for the observed Barkas effect.

\subsection{Origin of the Barkas Effect}
\label{sec:origin}

The solid curves in Fig.~\ref{fig:ept-vs-md} show the result of evaluating EPT for the velocity relaxation rate at the conditions of the MD simulations, $\nu_{ei}=\sqrt{2/3\pi}\Gamma_e^{3/2}\Xi_{ei}^{(1,1)}\omega_{pe}$, derived from the momentum moment of the collision operator (see Sec.~\ref{sec:v-relax} for more details).
Since the final value of the ion coupling strength in the MD simulations depends slightly on $\Gamma_e$ (and therefore not all points in the same frame have the same $\Gamma_i$), the EPT relaxation rates are shown as a band of values spanning the highest and lowest values that occurred in the simulations.
Across all coupling strengths investigated, the EPT predictions qualitatively reproduce the Barkas effect seen in the MD simulations.
Furthermore, for $\Gamma_e\lesssim 0.5$, EPT accurately predicts the numerical value of the relaxation rate.
At higher electron coupling strength, the approximation of Sec.~\ref{sec:model} used in the EPT calculations is no longer valid, so it is unsurprising that the model for the potential of mean force used here does not reproduce the MD relaxation rates in that regime.

\begin{figure*}
  \centering
  \includegraphics[width=\textwidth]{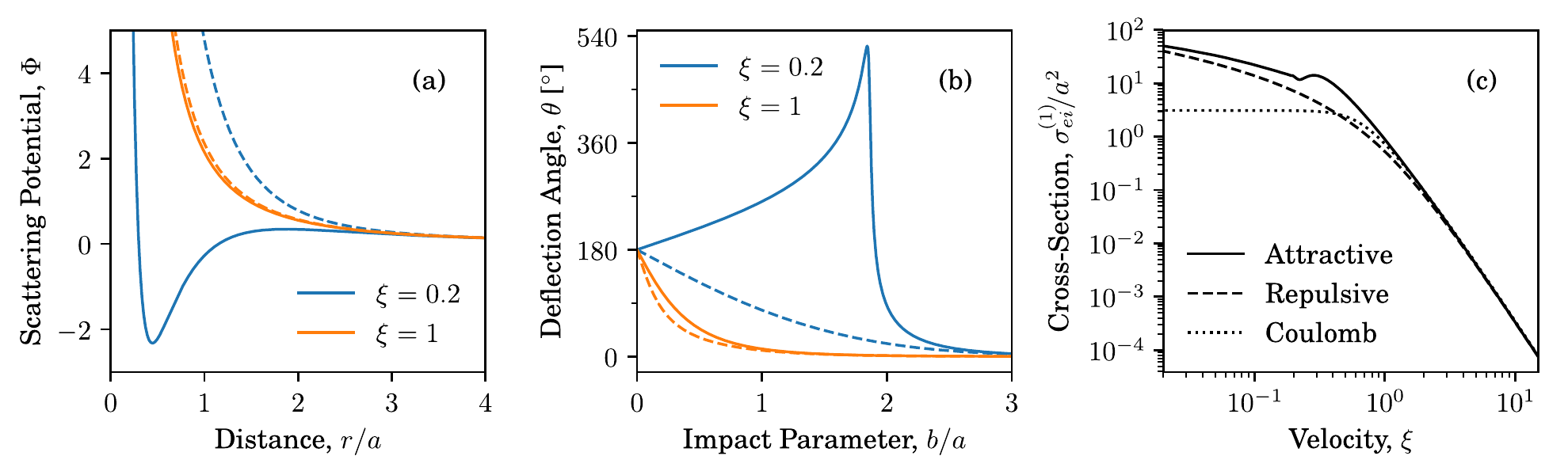}
  \caption{
    From left to right for a plasma with $\Gamma_i=1$ and $\Gamma_e=0.35$:
    (a) The scattering potential for binary collisions with impact parameter $b=1.5a$. 
    (b) The deflection angle, showing the onset of multiple-pass orbits for subthermal electrons.
    (c) The momentum-transfer cross-section. 
    In each, solid and dashed lines are for electron-ion and positron-ion, respectively.
    In (c), the dotted line is the cross-section for cut-off Coulomb collisions with $b_{\max}=\lambda_D$.
  }
  \label{fig:bigfig}
\end{figure*}
The accuracy of EPT in predicting the Barkas effect sheds light on its physical origin, namely that it can be understood in terms of the kinematics of attractive versus repulsive screened collisions.
The central dynamical quantity for understanding the Barkas effect is the scattering potential, $\Phi_{ei}(r,b,u) = b^2/r^2+2w_{ei}(r)/m_{ei}u^2$.
Figure~\ref{fig:bigfig} shows how the sign of $w_{ei}(r)$ alters the structure of the scattering potential and how this carries through to the deflection angle and cross-section.
For high-velocity collisions ($\xi\ge1$), the collision dynamics are controlled mainly by the centrifugal repulsion $b^2/r^2$, which is sign-independent.
The Barkas effect instead is prominent at lower relative velocities ($\xi<1$).
For positron-ion collisions, decreasing velocity causes $\Phi_{ei}(r)$ to smoothly transition from being centrifugally dominated to being dominated by $w_{ei}(r)$.
The cross-section likewise transitions from being dominated by small-angle scattering to large-angle scattering.
Electron-ion collisions, however, allow the possibility of spiral scattering when $\xi^2\lesssim Z\Gamma_e^{3/2}$ whereby the centrifugal and attractive terms of $\Phi_{ei}(r)$ compete at intermediate range to form a potential well.\cite{AlphaReview,KhrapakPRL2003}
This behavior is illustrated in the solid $\xi=0.2$ curve of Fig.~\ref{fig:bigfig}a.
As an electron traverses this well, it experiences a large angular deflection, sometimes involving multiple full rotations of the electron around the ion before eventually scattering away, as seen in the $\xi=0.2$ curve of Fig.~\ref{fig:bigfig}b.
The net effect is that electrons scatter through large angles over a wider range of impact parameters than do positrons, leading to a bigger electron-ion cross-section for momentum transfer compared to positron-ion collisions.
In turn, this leads to enhanced transport rates.

\subsection{Absence from Other Models}
\label{sec:absence}

Besides EPT, there are two main theoretical approaches to collisional transport in strongly coupled plasma: those that extend the Landau-Spitzer theory\cite{LeePF1984,LiPRL1993,DharmaWardanaPRE1998,GerickePRE2002,WhitePRE2017} and those that extend the Lenard-Balescu theory.\cite{DaligaultPRE2009,VorbergerPRE2010,benedict_pre_12,BenedictPRE2017}
Neither of these approaches predict the effect shown here.
The reason is that neither accurately models how screening influences close interactions.

Extensions of the Landau-Spitzer theory retain the idea that the binary Coulomb collision is the basic unit of transport.
That is, the deflection angle is given by the Rutherford formula $\theta_{ss'} = 2\arctan(q_sq_{s'}/m_{ss'}u^2b)$, so that the (first) momentum-transfer cross-section takes the form $\sigma^{(1)}_{ss'}(u)= 2\pi (q_sq_{s'}/m_{ss'}u^2)^2\ln[1 + (m_{ss'}u^2b_{\max}/q_sq_{s'})^2]$ after truncating the divergent integral in Eq.~\eqref{eq:sigma} at a characteristic interaction range $b_{\max}$.
It is in the choice of $b_{\max}$ that these approaches attempt to insert many-body screening physics, typically by interpolating between $\lambda_{D}$ for weak coupling and $a_i$ for strong coupling.\cite{LeePF1984,GerickePRE2002}
However, the resulting cross-section is independent of whether the collision is attractive or repulsive.
This is a peculiarity of inverse-square forces; the assumption of Coulomb collisions prevents Landau-Spitzer-based models from exhibiting a Barkas effect.

The lack of a Barkas effect in Landau-Spizter-based models is a symptom of not accounting for screening in the kinematics of close collisions.
The effect cannot be adequately recovered by means of altering $b_{\max}$.
Recall that the EPT model of Sec.~\ref{sec:model} was able to reproduce the Barkas effect observed in MD by changing only the sign --- not the range --- of the interaction potential.
While one could imagine prescribing a $b_{\max}$ that reproduces the Barkas effect, the underlying collision physics would be dubious.
This shortcoming of the Landau-Spitzer approach is evocative of the Salpeter enhancement of nuclear reaction rates, where screening between ions decreases the energy barrier for close collisions, a phenomenon also missed in a treatment based on impact parameter cutoffs.\cite{SalpeterAJP1954,AndereggPRL2009}

The Barkas effect is also not accurately treated in linear response theory, such as extensions of the Lenard-Balescu collision operator.
In these models, strong-coupling effects are included through static local field corrections (LFCs), $G_{ss'}(k)$, to the polarization potential around charges.
For instance, the electron-ion relaxation rate derived by Daligault and Dimonte takes the form (for classical plasmas)
\begin{subequations}
  \begin{equation}
    \begin{split}
      \nu_{ei} = \mathrm{const.}\times&\iint
      \left|\frac{\FT \phi_{ei}(k)}{D(k,\omega)}\right|^2 [1-G_{ei}(k)] \\
      &\quad \times\Im{\chi_e^0(k,\omega)}\Im{\chi_i^0(k,\omega)}\,d\omega d\vec k
    \end{split}
  \end{equation}
  \begin{equation}
    D(k,\omega) = \mathrm {det}\left\{\delta_{ss'} - \FT\phi_{ss'}(k)[1-G_{ss'}(k)]\chi_{s}^0(k,\omega)\right\}
  \end{equation}
\end{subequations}
where $\chi_s^0(k,\omega)$ is the free-particle density-density response function, and the function $\FT \phi_{ei}(k)/|D(k,\omega)|$ may be interpreted as a screened electron-ion interaction.\cite{DaligaultPRE2009}
(See Eqs.~(23-26) in Ref.~\onlinecite{DaligaultPRE2009} for comparison with related models.)
Since the electron-ion interaction only appears squared, a charge-sign asymmetry can only arise via a difference between electron-ion versus positron-ion LFCs.
That is, generalized Lenard-Balescu models predict a sign asymmetry due to the \emph{shape} of particles' dielectric dressing, but not the overall \emph{sign} of the interaction.
Physically, this is because approaches based on linear dielectric response model interactions via the correlation of linear fluctuations.
This excludes large-angle close collisions, which are a nonlinear phenomenon in this respect.

This basic limitation of linear response-based collision models becomes especially clear in comparison to the EPT model derived in Sec.~\ref{sec:model}.
The point of connection is the relationship between the direct correlation functions and the static LFCs,\cite{IchimaruBook,HansenMacDonald}
\begin{equation}
  \FT c_{ss'}(k) = -\frac{\FT\phi_{ss'}(k)}{k_BT_{ss'}}[1-G_{ss'}(k)]
  .
\end{equation}
The assumption of weak electron coupling in the present model (Sec.~\ref{sec:model}) translates directly to setting $G_{ei}(k)=G_{ee}(k)=0$.
Thus, to the same level of approximation used in the EPT computations shown in Fig.~\ref{fig:ept-vs-md}, the linear polarization potential approach does not predict a Barkas effect.
Such an approximation is necessary for making quantitative predictions because it is not known how to compute the LFCs of a classical electron-ion plasma for reasons described in Ref.~\onlinecite{DaligaultPRE2009} and in Appendix~\ref{sec:alpha-extrap}.
Even if one had a good model for $G_{ei}(k)$, it seems unlikely that using it in generalized Lenard-Balescu theory will accurately capture the Barkas effect, since it arises primarily due to the sign of the screened potential, not the shape.

\section{Impact on Transport Processes}
\label{sec:transport}

\subsection{Velocity Relaxation in UNPs}
\label{sec:v-relax}

In a uniform plasma, the electron drift velocity evolves according to $m_en_e\partial_t\vec V_e  = \vec R_{ei}$, where $\vec R_{ei}$ is the rate of change of the electron momentum density due to electron-ion collisions, i.e., the momentum moment of the collision operator.
For the Boltzmann collision operator, this may be evaluated as\cite{FerzigerKaper}
\begin{equation}
  \label{eq:R-general}
  \vec R_{ei} = -m_{ei}\iint u \vec u \sigma^{(1)}_{ei}(u) f_e(\vec v_e)f_i(\vec v_i)\,d\vec v_id\vec v_e
  ,
\end{equation}
where $\vec u=\vec v_e-\vec v_i$.
Eq.~\eqref{eq:R-general} is general for any form of the distribution functions, but when they are flow-shifted Maxwellians, $\vec R_{ei}$ takes the simple form
\begin{equation}
  \label{eq:R-max}
  \vec R_{ei} = -m_{ei}n_e\bar\nu_{ei}\,(\vec V_e-\vec V_i)
  ,
\end{equation}
where the velocity-dependent relaxation rate is
\begin{equation}
  \label{eq:nu-bar}
  \bar\nu_{ei}(\eta) = \frac{16\sqrt{\pi}Z^2e^4n_i}{3m_em_{ei}\bar v_{ei}^3}\bar\Xi_{ei}(\eta)
  ,
\end{equation}
in which $\eta = |\vec V_e-\vec V_i|/\bar v_{ei}$ is the dimensionless relative drift speed, $\bar\Xi_{ei}$ is a velocity-dependent generalization of $\Xi_{ei}^{(1,1)}$,
\begin{equation}
  \label{eq:Xi-bar}
  \bar\Xi_{ei}(\eta) = \frac{3}{32\eta^3}\int_0^\infty  \frac{\sigma_{ei}^{(1)}(\xi)}{\sigma_{0,ei}} \xi^2H(\xi,\eta)d\xi
  ,
\end{equation}
and $H(\xi,\eta)=(2\xi\eta+1)e^{-(\xi+\eta)^2}+(2\xi\eta-1)e^{-(\xi-\eta)^2}$.\cite{BaalrudAIP2016}
When the drift kinetic energy is small compared to the thermal energy ($\eta^2\ll1$),
\begin{equation}
  \label{eq:Xi-bar-expand}
  \bar\Xi_{ei}(\eta) = \Xi_{ei}^{(1,1)}\left[1 + c\eta^2 + O(\eta^4)\right]
  ,
\end{equation}
where $c=[6\Xi_{ei}^{(1,2)}-15\Xi_{ei}^{(1,1)}]/5\Xi_{ei}^{(1,1)}$.
Since ratios of generalized Coulomb logarithms are order-unity quantities\cite{BaalrudPOP2014} and the simulations in Sec.~\ref{sec:md} were performed at conditions such that $\eta\le 0.1$, the drift-velocity dependent corrections to the generalized Coulomb logarithm may be neglected here.
With the further simplifications that $m_e\ll m_i$, $T_e\ge T_i$, and $Zn_i=n_e$, a theoretical prediction for the velocity relaxation rate pertinent to the present MD simulations may be computed from
\begin{equation}
  \label{eq:nu}
  \bar\nu_{ei}(0) = \nu_{ei}
  \approx \sqrt{\frac{2}{3\pi}}Z\Gamma_e^{3/2}\Xi_{ei}^{(1,1)}\,\omega_{pe}
  .
\end{equation}
The EPT data shown in Fig.~\ref{fig:ept-vs-md} is the result of evaluating Eq.~\eqref{eq:nu} using the potential of mean force model described in Sec.~\ref{sec:model}.
\begin{figure}
  \centering
  \includegraphics[width=8.6cm]{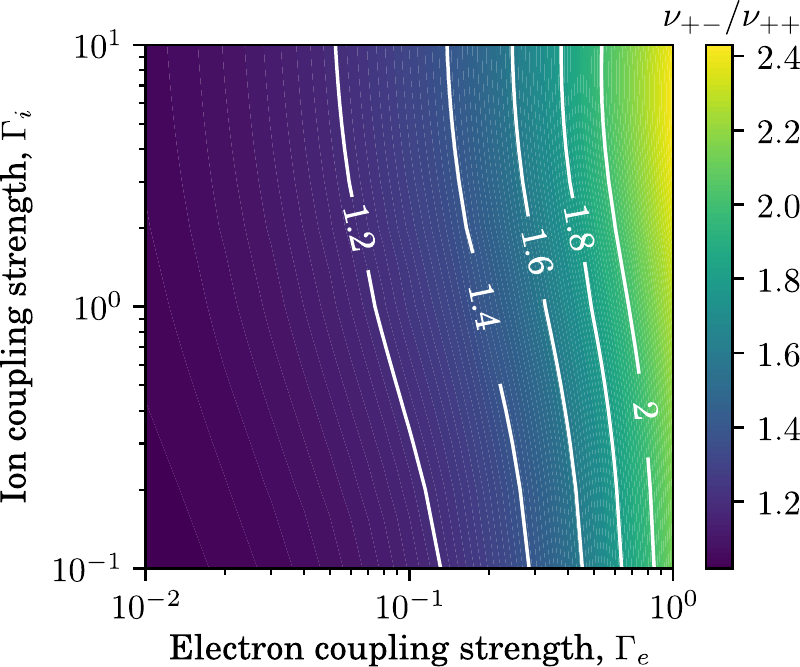}
  \caption{Ratio of the electron-ion to positron-ion collision frequency predicted by EPT as a function of the coupling strengths.}
  \label{fig:barkas}
\end{figure}

Velocity relaxation was the basic process measured in the experiments by Chen et al.\ on electron center-of-mass oscillation damping in UNPs.\cite{ChenPOP2016,ChenPRE2017}
Comparison of the measured damping rates with those obtained by hybrid MD-Monte Carlo simulations implementing various models for the electron-ion collision operator showed that the theories systematically underestimated the damping rate.
However, all the theories considered make either a weak-coupling approximation that precluded any Barkas effect or were based on models for strongly coupled plasmas with repulsive interactions.
Based on Fig.~\ref{fig:barkas}, though, at the strongly coupled conditions reported in those experiments, $\Gamma_e=0.35$, EPT predicts that the Barkas effect could contribute as much as a $50-70\%$ increase to the predicted collision rate compared to theories based on repulsive interactions, depending on the ion coupling strength.
If binary collisions were the dominant mode of damping in the experiments, then correcting for the sign of the electron-ion interaction could close the gap between collision models and measurements reported in Ref.~\onlinecite{ChenPRE2017}.
However, at the high $\Gamma_e$ they obtained, inelastic collisions between electrons and Rydberg atoms are also expected to influence the damping of the electron motion,\cite{RobertsPrivateComm,GuptaPRL2007} and it is not clear at present which is the dominant damping mechanism as $\Gamma_e$ increases.

\subsection{Temperature Relaxation in UNPs}
\label{sec:t-relax}

The temperature evolution in a uniform plasma is given by $\frac{3}{2}m_en_ek_B\partial_tT_e=Q_{ei}$, where $Q_{ei}$ is the rate of change of the electron internal energy density due to collisions, i.e., the $\frac{1}{2}m_e|\vec v_e-\vec V_e|^2$ moment of the collison operator.
For the Boltzmann collision operator,\cite{FerzigerKaper}
\begin{equation}
  \label{eq:Qei}
  \begin{split}
    Q_{ei} = -m_{ei}&\iint u\vec u\cdot\left(\vec v_i+\frac{m_{ei}}{m_i}\vec u\right)\sigma_{ei}^{(1)}(u)
    \\
    &\quad \times f_e(\vec v_e)f_i(\vec v_i)\,d\vec v_id\vec v_e
    -\vec V_e\cdot\vec R_{ei}
    .
  \end{split}
\end{equation}
Again, taking the distribution functions to be flow-shifted Maxwellians, one finds\cite{BaalrudAIP2016}
\begin{equation}
  \label{eq:Q-split}
  Q_{ei} = -3k_B\frac{m_{ei}}{m_i}n_e\tilde\nu_{ei}\,(T_e-T_i) - \frac{v_{Te}^2}{\bar v_{ei}^2}(\vec V_e-\vec V_i)\cdot\vec R_{ei}
  .
\end{equation}
The first term describes temperature relaxation between the electrons and ions at the velocity-dependent rate
\begin{equation}
  \label{eq:nu-tilde}
  \tilde\nu_{ei}(\eta) = \frac{16\sqrt\pi Z^2e^4n_i}{3m_em_{ei}\bar v_{ei}^3}\tilde\Xi_{ei}(\eta)
\end{equation}
which involves a different velocity-dependent Coulomb logarithm
\begin{equation}
  \label{eq:tilde-Xi}
  \tilde\Xi_{ei}(\eta) = \frac{1}{8\eta} \int_0^\infty \frac{\sigma_{ei}^{(1)}(\xi)}{\sigma_{0,ei}}\xi^4I(\xi,\eta)d\xi,
\end{equation}
with $I(\xi,\eta)=e^{-(\xi-\eta)^2}-e^{-(\xi+\eta)^2}$.\cite{BaalrudAIP2016}
For small relative drift speeds,
\begin{equation}
  \tilde\Xi_{ei}(\eta) = \Xi_{ei}^{(1,1)}[1 + c\eta^2 + O(\eta^4)],
\end{equation}
where $c=[2\Xi_{ei}^{(1,2)}-3\Xi_{ei}^{(1,1)}]/3\Xi_{ei}^{(1,1)}$ is again of order unity.
The second term in Eq.~\eqref{eq:Q-split} represents frictional heating; since it is $O(\eta^2)$ compared to the temperature relaxation, it is disregarded here.
Comparing Eq.~\eqref{eq:nu-tilde} with Eq.~\eqref{eq:nu-bar} for small drift velocity, it is evident that both the temperature and velocity relaxation depend on the same quantity $\bar\nu_{ei}(0) = \tilde\nu_{ei}(0) = \nu_{ei}$, given by Eq.~\eqref{eq:nu}.
Thus the Barkas effect influences the temperature relaxation rate in exactly the same way as shown in Fig.~\ref{fig:barkas} for the velocity relaxation rate.

\subsection{Electrical Conductivity in Dense, Thermal Plasmas}
\label{sec:cond}

The Barkas effect is also expected to extend to electron-ion transport in dense plasmas with partially Fermi-degenerate electrons.
Since EPT has not yet been formulated for quantum-mechanical transport, it would be incorrect to directly apply the results obtained so far for classical plasmas to dense plasmas.
Nevertheless, it is expected that the transport coefficients of dense plasmas are smooth functions of the electron degeneracy.\cite{DaligaultPRL2017,DaligaultPOP2018}
This implies that a Barkas effect should still be present in these systems at partial electron degeneracy.

In dense plasmas, theoretical models for electron transport are difficult to validate due to a lack of a first principles quantum simulation method.
An alternative approach is to treat the dynamics of the plasma classically but to model quantum-mechanical effects by altering the interaction potential between particles.
The main reason to do so is that it allows the use of classical MD to test theory.
A recent example is the semi-classical model of electrical conductivity presented by Whitley et al.,\cite{WhitleyCPP2015} who performed MD simulations of hydrogen plasmas using the Dunn-Broyles potential\cite{DunnPR1967}
\begin{equation}
  \label{eq:phi-db}
  \phi^{DB}_{ss'}(r) = \frac{q_sq_{s'}}{r}\left[1 - \exp\left(-\pi r/\lambda_{ss'}\right)\right]
\end{equation}
with $\lambda_{ss'}=\hbar/\sqrt{2m_{ss'}k_BT}$.
They compared their MD results with theoretical predictions based on the classical Lenard-Balescu collision operator.
Due to the use of the Lenard-Balescu collision operator, there is no charge-sign dependence in their model for the electrical conductivity.

For comparison, EPT predictions for the electrical conductivity were computed from the second-order Chapman-Enskog formula,
\begin{subequations}
  \begin{equation}
    \label{eq:sigma2}    
    [\sigma]_2 = \frac{[\sigma]_1}{1 - \Delta}
  \end{equation}
  \begin{equation}
    \label{eq:sigma1}
    [\sigma]_1 =
    \frac{3}{16\sqrt{\pi}}
    \frac{(2k_BT)^{3/2}}{e^2\sqrt{m_e} \Xi_{ei}^{(1,1)}}
  \end{equation}
  \begin{equation}
    \label{eq:delta}
    \Delta =
    \frac{(5\Xi_{ei}^{(1,1)} - 2\Xi_{ei}^{(1,2)})^2/\Xi_{ei}^{(1,1)}}
    {25\Xi_{ei}^{(1,1)} - 20\Xi_{ei}^{(1,2)} + 4\Xi_{ei}^{(1,3)} + 2\sqrt{2}\Xi_{ee}^{(2,2)}}
    ,
  \end{equation}
\end{subequations}
taken in the approximation $m_e\ll m_i$.\cite{FerzigerKaper}
The factor $(1-\Delta)^{-1}$ is the ``Spitzer'' correction accounting for skewness in the electron distribution function due the applied electric field.\cite{SpitzerPR1953}
The electron-electron and electron-ion Coulomb logarithms were computed using the potentials of mean force obtained from equilibrium HNC using the Dunn-Broyles potential above in Eqs.~\eqref{eq:oz} and~\eqref{eq:hnc-eq}.
That is, no weak-coupling approximations were made.
This was possible because the Dunn-Broyles potential is very soft as $r\to0$ for the densities and temperatures considered here, so these HNC calculations did not suffer from the numerical difficulties described in Sec.~\ref{sec:model} or Appendix~\ref{sec:alpha-extrap}.

\begin{figure}
  \centering
  \includegraphics[width=8.6cm]{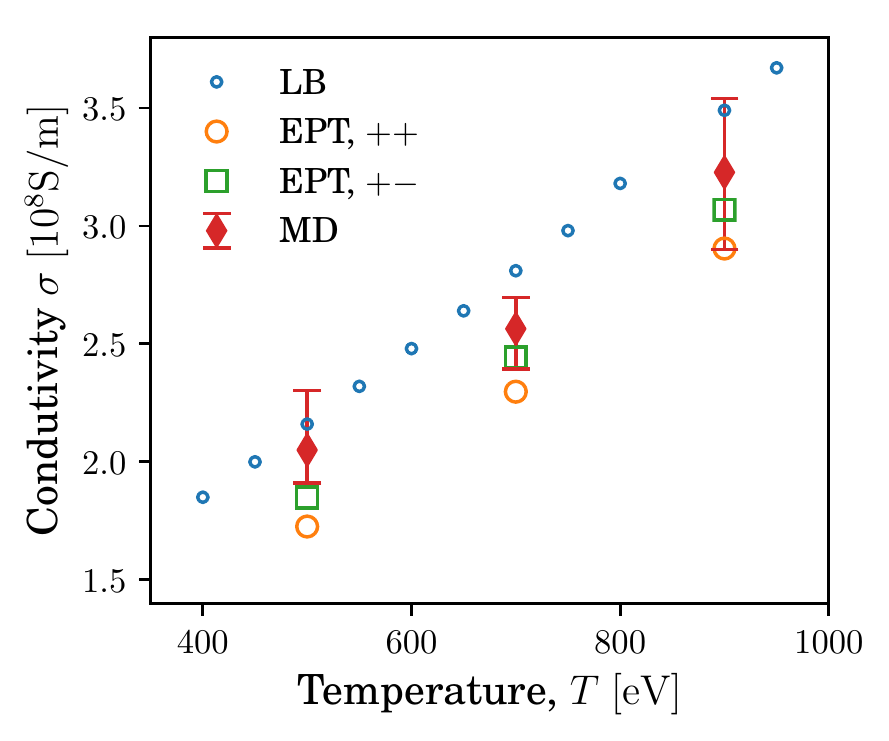}
  \caption{Semiclassical models of electrical conductivity of fully ionized hydrogen with density $\rho=40\mathrm{g/cm^3}$. Small blue circles: generalized Lenard-Balescu results from Ref.~\onlinecite{WhitleyCPP2015}. Orange circles: EPT calculations with positron-ion interactions. Green squares: EPT calculations with attractive electron-ion interactions. Red diamonds: MD results from Ref.~\onlinecite{WhitleyCPP2015}.}
  \label{fig:sigma}
\end{figure}

Figure~\ref{fig:sigma} compares EPT with the Lenard-Balsecu model and MD simulations of Ref.~\onlinecite{WhitleyCPP2015} for a fully ionized hydrogen plasma with mass density $\rho=40\mathrm{g/cm^3}$ and $k_BT=500,700,900\mathrm{eV}$.
Two sets of EPT results are shown; they differ only in the overall sign of the electron-ion potential of mean force.
At these conditions, the plasma is moderately coupled ($\Gamma=e^2/ak_BT\sim 0.1$) and the electron-ion and electron-electron interactions are quite soft ($\lambda_{ei}\sim 2a_e$).
Both the Lenard-Balescu model and attractive EPT offer reasonable predictions for the conductivity in light of the large variability in the MD results.
However, in comparing the attractive and repulsive EPT calculations, it is seen that the repulsive interaction leads to a conductivity that is systematically about $10\%$ smaller.
That is, softening the Coulomb interaction results in attractive interactions having a \emph{smaller} collision rate compared to repulsive ones.
This means that within this simplistic model, not only does a charge-sign asymmetry effect persist, but in fact it inverts compared to the classical case.
This is in part because in the Dunn-Broyles interaction the large-angle and spiral scattering events described in Sec.~\ref{sec:origin} onset at lower velocities compared to the attractive Coulomb interaction, so they do not contribute significantly to transport at the conditions considered here.
It remains to be seen if a more accurate treatment of quantum-mechanical effective-potential scattering would exhibit a similar inversion of the Barkas effect.

\section{Conclusions}
\label{sec:conclusions}

It has been demonstrated that the transport rates of plasmas exhibit a dependence on the sign of the electron charge analogous to the Barkas effect in charged particle stopping.
This is an emergent consequence of strong coupling that arises when one makes a detailed account of how screening in the plasma alters the trajectories of binary encounters between electrons and ions; the effect cannot be recovered by treating collisions as Coulomb interactions with truncated range, nor is it captured by the leading theories of classical transport in strongly coupled plasma based on linear response.
The effect is small at weak electron coupling where these approaches still approximately hold, but as the electron coupling strength approaches unity, the electron-ion collision rate is significantly enhanced compared to that for positron-ion collisions.
UNP experiments may be able to measure this fundamental physical effect in plasma transport.
It is also expected to influence essentially all neutral plasmas at strong coupling, including dense plasmas (though electron degeneracy must also be accounted for to accurately treat these systems).
\section*{Acknowledgements}
\label{sec:ack}

We wish to thank J\'er\^ome Daligault and Jacob Roberts for stimulating conversations, as well as Lorin Benedict for sharing the LB and MD electrical conductivity data shown in Fig.~\ref{fig:sigma}.
This material is based upon work supported by the National Science Foundation under Grant No.~PHY-1453736. It used the Extreme Science and Engineering Discovery Environment (XSEDE), which is supported by NSF Grant No.~ACI-1053575 under Project Award No.~PHY-150018.

\appendix

\section{Fitting and Error Estimation of MD Results}
\label{sec:fit}

Because the drift velocity was necessarily small compared to the electron/positron thermal speed, each individual run contained significant fluctuations in $\vec V_e(t)$.
Thus, fitting individual drift-velocity time series was inappropriate.
Instead, the sample mean and standard deviation of the drift velocity were used to obtain the relaxation.
The ``sample'' here means that the mean and standard deviation were taken over 100 independent replicas of a single combination of $\Gamma_i^0$ and $\Gamma_e^0$.
Specifically, the relaxation rate was determined by minimizing~\cite{NumericalRecipes}
\begin{equation}
  \chi^2 = \sum_{i=1}^{N_t} 
  \frac{
    |\bar V(t_i) - V_0 \exp[-\nu_{ei}(t_i-t_{\mathrm{kick}})]|^2
  }{
    \sigma^2(t_i)
  }
\end{equation}
with respect to the fit parameters $V_0$ and $\nu_{ei}$, where $N_t$ is the number of time steps in the fitting interval,
\begin{equation}
  \bar V(t) = \frac{1}{N_{r}}\sum_{i=1}^{N_r} \hat z\cdot\vec V_{e,i}(t)
\end{equation}
is the sample mean drift velocity across $N_r$ runs, and
\begin{equation}
  \sigma^2(t) = \frac{1}{N_r}\sum_{i=1}^{N_r}[\hat z\cdot\vec V_{e,i}(t) - \bar V(t)]^2
\end{equation}
is the sample variance.
Confidence intervals for the fit parameters were taken from the diagonals entries of the fit parameter covariance matrix; the off-diagonal entries were always negligible in comparison.
The uncertainties in the fit parameters were typically very small (parts per hundred), as a result of the fit being highly over-constrained (thousands of time steps to determine two fit parameters).

The start of the time interval used to fit the drift velocity was chosen to be late enough after the kick that the non-exponential transient response was excluded.
Ideally, the fit should be started as late as possible, just not so late that the drift velocity falls below the thermal noise floor.
However, in the electron-ion systems, the competing requirement that the total run time be short constrains how late the fitting window can start.
It was found that starting $5\omega_{pe}^{-1}$ after the kick was suitable.
This was long enough after the kick that non-exponential decay was not apparent, yet still early enough that a substantial time was included ($10\omega_{pe}^{-1}$).
For the longer positron-ion simulations, there was much more flexibility in choosing the fitting interval.
The start was chosen to be $10\omega_{pe}^{-1}$ after the kick, and the end was chosen to be either the end of the simulation or the time at which the drift velocity had decayed by three $e$-foldings, whichever occurred earlier.
The coupling strengths associated with the fitted relaxation rate were the average $\Gamma_e$ and $\Gamma_i$ over the fitting interval.
Conservative estimates for the ``error'' in the coupling strengths were computed from the minimum and maximum values of $\Gamma_e$ and $\Gamma_i$ over the whole fit interval and across all included runs.

\section{Numerical Details of Softened Electron-Ion HNC Calculations}
\label{sec:alpha-extrap}

The HNC calculations presented in this paper were done using Fozzie, a free and open-source program to numerically solve the OZ or SVT equations.\cite{Fozzie}
The solution method is an iterative scheme in the same vein as Refs.~\onlinecite{SpringerJCP1973,NgJCP1974}.
The problem of solving the OZ or SVT equations subject to the HNC closure is cast as a fixed-point problem for the direct correlation functions, $\mathcal A[\tens c(r)] = \tens c(r)$, where $\tens c$ is the two-by-two matrix whose elements are $c_{ss'}$, and the operator $\mathcal A$ represents a sequence of steps: (1) Fourier transforming each $c_{ss'}(r)\to\FT c_{ss'}(k)$, (2) solving the OZ or SVT relations for the indirect correlation functions, $\FT\gamma_{ss'}(k)\equiv\FT h_{ss'}(k)-\FT c_{ss'}(k)$, (3) inverse Fourier transforming each $\FT\gamma_{ss'}(k)\to\gamma_{ss'}(r)$, and (4) evaluating $\exp(-\phi_{ss'}/k_BT_{ss'}+\gamma_{ss'}) - \gamma_{ss'} - 1$.
Given a trial solution $\tens c^{(i)}$, the application of $\mathcal A$ produces a new trial solution $\tens c^{(i+1)}=\mathcal A[\tens c_i]$.
This trial solution was then linearly mixed with the previous one: $\tens c^{(i+1)}\leftarrow \zeta \tens c^{(i+1)} + (1-\zeta) \tens c^{(i)}$, where the mixing parameter $\zeta$ was set a priori.
Self-consistency was judged by computing $\delta^{(i)}= \sqrt{\int_0^\infty\frac{1}{4}\sum_{s,s'=1}^2|c_{ss'}^{(i+1)}-c_{ss'}^{(i)}|^2 \,dr}$.
If $\delta^{(i)}<10^{-8}$, then the calculation was ended and $\tens c^{(i+1)}$ was taken to be the solution.
Otherwise, a new trial solution was generated from $\tens c^{(i+1)}$.

The rate of convergence depended both on the initial trial solution $\tens c^{(0)}$ and the mixing parameter $\zeta$.
When the system to solve had strong attractive interactions (as in the case $\alpha\to0$), the iteration did not converge at all unless $\tens c^{(0)}$ was already close to a solution and the iterative refinement was forced to progress slowly by setting $\zeta\ll1$.
To obtain HNC solutions with the smallest feasible values of $\alpha$, a sequence of solutions were generated, each with a smaller value of $\alpha$ than its predecessor.
To aid convergence, each run was initialized with the converged result of the preceding one.
If a run failed to converge, it was retried with a value of $\alpha$ closer to the previous run's and a more conservative choice of the mixing parameter $\zeta$.
This continued either until 50 attempts had been made or until the values of $\alpha$ between two successive attempts differed by less than $0.05\%$.
It was observed that this routine tended to terminate at $\alpha/a_e\approx \Gamma_e/10$.
The resulting potentials of mean force with those from MD showed unacceptable non-Coulombic behavior when $\Gamma_e\sim 0.1-1$, even when $r\sim a (\gg \alpha)$.


%

\bibliography{refs}

\end{document}